\documentstyle[preprint,revtex]{aps}
\math-with-secnums
\begin{document}
\draft
\def\doublespaced{\baselineskip=1.0\normalbaselineskip}
\let\doublespace = \doublespaced
\doublespace
\hfill{UM-P-93/01}

\hfill{OZ-93/01}

\begin{title}
Ten Dimensional $SO(10)$ G.U.T. Models \\
with Dynamical Symmetry Breaking
\end{title}
\author{B. E. Hanlon and G. C. Joshi}
\begin{instit}
Research Centre for High Energy Physics \\
School of Physics \\
University of Melbourne \\
Parkville, Victoria 3052 \\
Australia
\end{instit}
\begin{abstract}
We discuss the derivation of $SO(10)$ G.U.T. models from higher
dimensional theories with intermediate breaking scales. We then
present models based on the Coset Space Dimensional Reduction Scheme
with intermediate symmetry breaking induced by four fermion
condensates.
\end{abstract}
\newpage
\section{Introduction}
The phenomenological advantages of G.U.T. models with quark-lepton
unification
are well known. For example, the proton
decay rate may be suppressed. Also, the existence of a right
handed neutrino may account for the missing mass problem in cosmology
and provide a mechanism for a light mass left handed neutrino to exist
consistently in nature. G.U.T. models of this kind, such as $SO(10)$
or $SU(16)$, imply the existence of an intermediate mass scale
providing greater freedom to incorporate other phenomenological
features, such as a reasonable value for ${\sin}^{2}\theta_{w}${\cite
a}.

The concept of grand unification can be extended into the broader
framework of superstring theory{\cite b}
and superstring inspired models. These
models incorporate the notion of an extended space-time and
dimensional reduction to the four dimensions observed in
nature. Much work has been done in this field, in particular on models
which realise $E_{6}$ G.U.T. models in four dimensions\cite{b,c}.
These examples arise most naturally due to the properties of the
underlying space-time manifold and the relative ease with which a
symmetry breaking pattern to low energy physics can be incorporated,
such as with symmetry breaking by Wilson lines\cite{d,e}
. This is not the case
with $SO(10)$ models (we do not consider $SU(16)$ so that we do not
have to assume the existence of mirror families to cancel anomalies).
The existence of an intermediate symmetry (in particular incorporating
$SU(4)$ of colour) is not easily accomodated within such a
framework\cite{e,f}.
Interestingly, exploiting these extended symmetry breaking patterns by
the assumed existence of nonrenormalizable higher order operators
arising from spontaneous compactifications, can allow significant
phenomenology to appear at low energies{\cite g}.
For instance, rare kaon
decays could be accessable at current machine energies and so be a
signature of left-right symmetric models. These conjectures are
supported by only one particular example of reduction from a higher
dimensional theory, formulated by Wetterich{\cite h}.
His approach generates
chiral fermions by the dimensional reduction onto a non-compact
internal manifold of finite volume. Starting from an eighteen
dimensional theory, a six dimensional model emerges with an $SO(12)$
gauge symmetry. Considered as a gauge theory on $M^{4} \otimes S^{2}$,
the model is reduced to a four dimensional $SO(10)$ G.U.T. . Higgs
fields required for symmetry breaking are introduced into the model
from six dimensional $SO(12)$ representations which exhibit nonzero
coupling to the fermions.

While important in its own right, it would be useful to have an
alternate description of a higher dimensional model with intermediate
symmetry breaking. Within string inspired models this would be
particularly important due to the naturalness with which $SO(10)$
gauge models emerge with the appropriate fermionic representations
{\cite b}.
$SO(10)$ models inspired from $E_{6}$ gauge theories,
arising from manifolds with $SU(3)$ holonomy, have been
considered but contain a number of fermions within the $\bf 27$ of
$E_{6}$ which are not realized in nature and represent a compromise
solution to finding realistic $SO(10)$ models on appropriate
manifolds{\cite i}.
It has also been
demonstrated that the existence of a low energy supersymmetry breaking
could solve many phenomenological problems but a mechanism to
implement this remains speculative{\cite j}.

A particular string inspired approach to model building, which has been
applied with some success to $SU(5)$ G.U.T. models, is to impose
space-time invariance conditions on all the fields, known as Coset
Space Dimensional Reduction (CSDR)\cite{k,l}.
This has the advantage of
producing a finite number of states in four dimensions, as opposed to
an infinite tower of states in the harmonic expansion approach usually
employed{\cite m}, as well as providing a possible origin for the
Higgs mechanism.
To date, considerations on $SO(10)$ models within CSDR have
been ``diagonalized" to the standard model or rely upon imaginative
applications of Wilson lines
so as to avoid the problem
of the nonexistence of an intermediate Higgs mechanism\cite{l,n}.
However, there
is an alternative approach involving four fermion condensates,
breaking symmetries by a dynamical mechanism{\cite o}.
Indeed, dynamical
symmetry breaking has been the direction taken in some $SU(5)$ models
within this framework in order to avoid the problems of electroweak
symmetry breaking at the compactification scale\cite{n,p}.
In this paper
we will present realistic models which utilize this
mechanism. We will show that the
appropriate fermionic representations can emerge from CSDR and we will
discuss the construction of such condensates within the constraints of
this scheme. By introducing discrete symmetries onto the internal
manifold we can produce strong breaking of the $SO(10)$
G.U.T. and, more importantly, eliminate Higgs fields of geometrical
origin.

\section{Coset Space Dimensional Reduction and SO(10)}

In models constructed on extended space-times, gauge fields are
introduced as a possible origin for the Higgs mechanism and also
because they provide for the existence of low mass flavour chiral
fermions in four dimensions by having non-trivial field configurations
on the internal manifold\cite{m,q}.
While not in the spirit of a purely
gravitational model, this approach allows for an interesting new
approach to G.U.T.s by beginning with a larger, more generalized
symmetry in higher dimensions. Clearly, this is motivated by the
emergence of gauge symmetries as a natural part of super string
theory{\cite b}.

The space-time manifold in such a scheme is presumed to have the form
$M^{4} \otimes S/R$, where $M^{4}$ is four dimensional Minkowski space
and $S/R$ is a compact coset space. Rather than set to zero the field
dependance on the internal coordinates on $S/R$, CSDR provides a means
by which a field dependence can be maintained. The number of
space-time dimensions can be consistently reduced by imposing $S$
invariance on all the fields, so producing a finite number of fields
in four dimensions. That is, transformations under symmetries of $S/R$
are compensated by gauge transformations. Starting from a principal
fibre bundle with bundle group $G$ defined over $S/R$,
$S$ invariant connections
are characterized by linear maps from the
Lie algebra of $S$ to the Lie algebra of $G$ such that $\Phi_{i} : R
\rightarrow G$ is a faithful homomorphism
. Corresponding to such an
embedding, the gauge fields carrying vector indicies corresponding to
the additional internal dimensions,
which behave as scalar fields under four dimensional space-time
transformations, transform under $R$ as a vector $\bf v$ specified by
the embedding
\begin{equation}
adj(S) = adj(R) + {\bf v} \;\; .
\end{equation}
The $\Phi_{i}$, where $i$ corresponds to a generator of $\bf v$, then
satisfy the linear constraint condition
\begin{equation}
[ \Phi_{i} , \Phi_{j} ] = f_{ijk} \Phi_{k}\;\;
\forall j \in \Im (adj(R))
\;\; ,
\end{equation}
where $f_{ijk}$ are the structure constants of $S$,
and have arbitrary values. When $i$ corresponds to a generator of $R$
the $\Phi_{i}$ are not arbitrary and define a nontrivial $R$ bundle
over $S/R$. The gauge symmetry which survives this procedure , $H$, is
the centralizer of the image of $R$ in $G$.

It is found, by exploiting Schur's lemma, that an
unconstrained scalar field is obtained whenever the tensor product of
an induced representation of $R$ over $S/R$ and a representation of
$R$ in the adjoint of $G$ contains a singlet. Similarly, the surviving
fermionic fields in four dimensions are found by applying Schur's
lemma, this time with consideration to the branching rule of the
spinor representation of the coset space tangent group under $R$.
Starting from a vectorlike representation, flavour chirality in four
dimensions demands $Rank S = Rank  R${\cite r}.
This places severe restrictions
on the allowed coset spaces. Imposing the Weyl and Majorana conditions
futher requires that the the total space-time dimensionality be $D=2 +
8n$ when the fermionic representations are real. Thus the smallest
dimension from
which we can construct a model within this scheme is ten.

The scalar fields which emerge, identified with Higgs fields, form a
potential in the effective four dimensional theory from the relevant
terms in the ten dimensional gauge kinetic action. Importantly, it is
found that if $S$ has an isomorphic image in $G$ then the four
dimensional symmetry group, $H$, breaks to $K$, the centralizer of $S$
in $G$. This result is independant of whether the coset space is
nonsymmetric, in which case it is otherwise possible to manipulate
the radial parameters so as to have a Higgs potential with vanishing
order parameter\cite{ab}.
It turns out that in such cases all fermionic fields become
massive, at the order of the compactification scale, after symmetry
breaking. Thus, such models are not phenomenologically viable unless
these fields can be otherwise eliminated.

For convenience, we list the six dimensional coset spaces with $Rank S
= Rank R${\cite s}:
\hspace{4mm} $ SO(7) / SO(6) \: ;
 SU(4)/ {SU(3)\otimes U(1)} \: ;
 SP(4) / {(SU(2)\otimes U(1))}_{max} \: ;  $
\hspace{5mm} $SP(4) / {(SU(2)\otimes U(1))}_{nonmax} \: ; $
 $G_{2} / SU(3) \: ; $
${SP(4)\otimes SU(2)} / {SU(2) \otimes SU(2) \otimes U(1)} \: ; $
  $ {SU(2) \otimes SU(2) \otimes SU(2)} /
{U(1) \otimes U(1) \otimes U(1)} \: ;
 {SU(3)\otimes SU(2)} / {SU(2)\otimes U(1) \otimes U(1)} \: ; $
and
$ SU(3) / {U(1)\otimes U(1)} \; $.

As we have mentioned, CSDR has been applied with some success to
generating $SU(5)$ G.U.T. models in four dimensions\cite{t,u,v}
. From the set of
allowed coset spaces, it is clear that if we wish to have an $SO(10)$
model after dimensional reduction then we must consider gauge fields
in ten dimensions with rank at least seven. Furthermore, we choose not
to allow horizontal flavour symmetries to emerge. This greatly
restricts the range of groups that we can consider. For instance,
unitary groups have not played a large role in unified model building
within CSDR. In fact, only a $G=SU(8)$ model dimensionally reduced on
the manifold $G_{2}/SU(3)$, yielding $H=SU(5)\otimes U(1)$, has been
considered{\cite p}.
Beside
the symplectic groups, this leaves $E_{8}$ and $E_{7}$. By considering
$E_{8}$ we are implicitly including $SO(16)$ and $SO(17)$, the
orthogonal groups of the correct rank which also have real spinorial
representations, via the maximal embeddings{\cite w}
\begin{equation}
E_{8} \supset SO(16) \;\;\;\;\;\; SO(17)\supset SO(16) \;\; .
\end{equation}
Indeed, it is sometimes more convenient to consider, for example,
$SO(16)$ rather than $E_{8}$ since the next smallest irreducible
representation to the fundamental $\bf 248$ of $E_{8}$ is the $\bf
3875$. By considering $SO(16)$ instead, there is a greater range of
choice of starting representations in ten dimensions. Furthermore, in
the particular case of $SO(16)$, this allows the scalar and fermionic
representations to be seperated, i.e. breaking supersymmetry by the
initial boundary conditions. This can be useful when discrete
symmetries are introduced and the transformation properties of the
vector and spinor under such a symmetry are considered{\cite v}.

\section{Intermediate Symmetry and \\
 the Wilson Flux Breaking Mechanism}

The non-trivial
Higgs scalars which arise from CSDR have $|\Delta I_{w}|= 1/2$
breaking components but are not sufficient to break the four
dimensional G.U.T. group. Strong breaking of $H$ can be induced,
however, by the Wilson flux breaking mechanism\cite{d,e}.
In this
scheme, rather than consider $M^{4}\otimes B_{0}$ where $B_{0}={S\big/
R}$ is a simply connected manifold, we consider a gauge theory on
$M^{4}\otimes B$ with $B={B_{0}\big/ K^{S / R}}$ where $K^{S /
R}$ is a freely acting symmetry on $B_{0}$.( A group $K$ acts freely
on $B_{0}$ if for any element $k\in K$ other than the identity, the
equation $k(y)=y$ has no solution for $y\in B_{0}$ ). The space $B$ is
not simply connected with its fundamental group $\pi_{1}$ isomorphic
to $K^{S / R}$.
This means that there will be contours not contractable to a
point in the manifold. The unbroken gauge group which results turns
out to be the centralizer of the homomorphic image, $K^{H}$, of
$K^{S / R}$ in $H$. Furthermore it is found that the matter fields
which survive have to be invariant under the diagonal sum
\[ K^{S / R}\oplus K^{H} \;\; . \]
This mechanism is related to the Aharanhov-Bohm effect in
electrodynamics. The freely acting discrete groups on all possible six
dimensional coset spaces satisfying $rankR=rankS$ have
already been derived
. These fall into two classes corresponding to the
centre of $S$ and $W={W_{S}\big/ W_{R}}$ where $W_{S}$ and $W_{R}$ are
the Weyl groups of $S$ and $R$ respectively{\cite n}.
Under $W$, the $S/R$
vector and spinor have non-trivial transformation properties. By
appropriately embedding in $H$ it becomes possible that eigenstates of
fields will not be invariant under the gauge group and so are
eliminated from the model. In particular, this provides a way to
eliminate Higgs fields of geometrical origin. On symmetric coset
spaces, for which symmetry breaking is guaranteed to occur at the
compactification scale, this becomes
crucial{\cite r}.
Since we wish to produce symmetry breaking by fermionic
condensates we will always require that the Higgs fields vanish. In
this way we avoid the problem that $S$ may have an isomorphic image in
$G$.

Rather than breaking symmetries dynamically,
there is an alternate way in which intermediate scale symmetry
breaking may be introduced. This approach is more closely related to
that of Wetterich{\cite h}. We could introduce fundamental Higgs fields
transforming in particular representations of the original gauge group
$G$. Those components which survive in four dimensions, transforming
under $H$, must be $R$ singlets. Note, however, that most of the
allowed coset spaces contain $U(1)$ factors in $R$. Being abelian,
these cannot be ``centralized away". Although these factors can be
essentially omitted by setting the coupling strengths to zero since
they receive seperate renormalization from the others{\cite x},
they will
affect the allowed couplings. This is just a statement of gauge
invariance under $G$. Fermionic fields derived from such manifolds
will carry non-trivial $U(1)$ quantum numbers.
The relevant tensor product of two such fields with a
Higgs field with zero $U(1)$ charge will not produce a gauge singlet.

If we instead consider manifolds without $U(1)$ factors in $R$, only
$S^{6} \simeq SO(7)/SO(6)$ emerges as a candidate ($G_{2}/SU(3)$ would
yield an $SO(10)$ model with an additional, unwanted, $U(1)$ gauge
symmetry which cannot be omitted).
Indeed, this was the the only example where appropriate
$SO(10)$ Higgs fields arose as $R$ singlets, particularly the $\bf
126$ of $SO(10)$. It is, also, on this simplest example of an internal
manifold that the naturalness with which
$SO(10)$ models emerge can be demonstrated. Starting from an $E_{8}$
theory, the $R=SO(6)$ group is identified with the subgroup appearing
in the decomposition\cite{l,n}
\[ E_{8} \supset {\underbrace{SO(6)}_{R}} \otimes SO(10) \;\; , \]
\begin{equation}
\bf
248 = (15,1) + (1,45) + (6,10) + (4,16) + ({\overline 4}, {\overline
16}) \;\; .
\label{decomp1}
\end{equation}
The surviving four dimensional gauge group will then be $H =
C_{E_{8}}(SO(16))=SO(10)$. The $SO(6)$ content of the $S^{6}$ vector
and spinor are $\bf 6$ and $\bf 4$ respectively{\cite r}.
Comparing this with
(\ref{decomp1}), the surviving Higgs fields transform as a $\bf 10$ of
$SO(10)$ and the left handed fermions as a $\bf 16$, if we choose the
ten dimensional theory to be supersymmetric. Being geometrical
in origin, with an order parameter associated with the
compactification scale, the $\bf 10$ will produce electroweak symmetry
breaking at a phenomenologically unacceptable large energy. Turning to
the Wilson flux breaking mechanism, we note that this manifold has a
$Z_{2}$ discrete symmetry in $W${\cite n}.
Under this discrete symmetry the
$SO(6)$ vector and spinor transform as:
\[
\rm{vector}
\left \{
\begin{array}{c}
{\bf 6} \leftrightarrow {\bf 6}
\end{array}
\right .
\;\;\;\;
\]
\begin{equation}
\rm{spinor}
\left \{
\begin{array}{c}
{\bf 4} \leftrightarrow {\bf {\overline{4}}}
\end{array}
\right .
\;\; .
\end{equation}
Thus, the Higgs field is unaffected and so survives. So while it
is ``very satisfying" that $SO(10)$ with the
correct fermion representation emerges in a natural way we cannot
rely upon this model.

Diagonalization to the standard model for this example has been
performed by embedding $Z_{2}^{S/R}$ into a discrete subgroup of the
$U(1)$ appearing in the decomposition{\cite n}
\begin{equation}
SO(10) \supset SU(2)\otimes SU(2)\otimes SU(4) \;\; ; \;\; SU(4)\supset
SU(3)\otimes U(1) \;\; .
\label{decomp2}
\end{equation}
The Wilson flux mechanism breaks the four
dimensional gauge group $H = C_{E_{8}}(SO(6))=SO(10)$ down to $H' =
C_{E_{8}}(SO(6)\otimes Z_{2}^{H}) = SU(3)\otimes SU(2)\otimes
SU(2)\otimes U(1)$. Since $S=SO(7)$ has an isomorphic image in $G$ the
symmetry breaking of $H$ by the geometrical Higgs is known: $K' =
C_{E_{8}}(SO(7))=SO(9)$. Both symmetry breaking mechanisms acting
together give a final unbroken gauge group $K = K' \cap H' =
SU(2)\otimes U(1)\otimes SU(3)$, where the $SU(2)$ in $K$ is the
diagonal sum of the previous $SU(2)$'s.

Alternatively a more ambitious embedding of discrete symmetries may be
pursued{\cite l}
. For example, suppose $S^{6}$  is divided out by $(Z_{2}\times
Z_{2})^{S/R}$ where one $Z_{2}$ is the centre of $SO(7)$ and the other
is in $W$. One $Z_{2}^{S/R}$ is identified with a $Z_{2}$ subgroup of
the $U(1)$ appearing in the decomposition $SO(10)\supset SU(5)\otimes
U(1)$, for which the Higgs and fermions have the branching rule
\begin{eqnarray}
{\bf 10} & = & {\bf 5}(2) + {\overline{\bf 5}}(-2) \nonumber \\
{\bf 16} & = & {\bf 1}(-5) + {\overline{\bf 5}}(3) + {\bf 10}(-1) \;\;
{}.
\end{eqnarray}
The $Z_{2}$ subgroup is chosen to be $Z_{2}={\rm exp}[i(n+1)\pi]$, $n$
being the $U(1)$ quantum number. The second $Z_{2}^{S/R}$ is
embedded in a $Z_{2}$ subgroup of the hypercharge under $SU(5)\supset
SU(2)\otimes SU(3)\otimes U(1)$, such that all the components of the
fundamental representation are invariant. The result is a model
yielding the standard model in four dimensions with a family of
fermions from the ${\bf 16}$ of $SO(10)$ and no surviving scalars.
Electroweak symmetry breaking must now rely upon dynamical means in
both cases.

Both these approaches attempt to give realistic models
in the absence of an intermediate Higgs mechanism. Clearly many
other similar examples could be constructed depending on the choice of
embedding for the discrete symmetries.

As with this last example we could try embedding $Z_{2}^{S/R}$ into
the $U(1)$ subgroup appearing in (3.3) in such a way as to yield an
$SU(2)\otimes SU(2)\otimes SU(3)\otimes U(1)$ model in four dimensions
where the ${\bf 10}$ is odd under $Z_{2}^{H}$. However, we note that
the bidoublet component of the ${\bf 10}$ is a singlet under this
$U(1)$ and so cannot be odd. Consequently we are still unable to
eliminate this field. Any considerations on eliminating this Higgs
field therefore rest with $Z_{2}^{S/R}$.

It should be pointed out that, even if we start with a supersymmetric
model in ten dimensions, supersymmetry will never survive to low
energies within CSDR. On symmetric coset spaces this is because the
constraints explicitly break $N=1$ supersymmetry{\cite r}. While the
constraints on nonsymmetric coset spaces preserve $N=1$
supersymmetry, there exists a purely geometric term which emerges from
the ten dimensional fermionic kinetic action, written as $V${\cite y}.
The
non-vanishing matrix elements of $V$ correspond to $R$ singlets, so
that gaugino fields in four dimensions acquire superheavy masses. It
may be possible to overcome this by introducing torsion onto
nonsymmetric cosets but this has yet to be demonstrated{\cite l}.
This, then, rules out implementing the generalized notion of the
see-saw mechanism where the neutrino mass problem can be tackled by
assuming a non zero vacuum expectation value for the scalar
superpartner of the right handed neutrino{\cite j}.

\section{Models with Dynamical Symmetry Breaking}

It is known from lattice calculations that it is possible to
generalize the Higgs phenomenon to a dynamical symmetry breaking
scheme, described in a gauge invariant way{\cite z}
. The existence of a gauge
symmetry breaking potential is associated with four fermion
condensates such that
\begin{equation}
< {\cal C} > \neq 0 \;\; ,
\end{equation}
where $\cal C$ is a four fermion gauge singlet operator. Four fermion
condensates are considered since for chiral gauge theories a quadratic
mass condensate does not exist.
It can be
shown, under a set of general assumptions, that an anomaly free set of
fermions, including exotics, which are capable of forming such
condensates, has the form{\cite o}
\begin{equation}
{\bf R_{L} = n.16 + 144}\;\;\; {\rm where} \;\;\; {\bf n} = 2,3,4
\;\;,
\end{equation}
for an $SO(10)$ model.
Clearly, two forms of gauge singlet condensates can be constructed:
\begin{equation}
<{\cal C} > = <LLLL> \;\;\; {\rm or}
 \;\;\;  <{\cal C} > = <LL{\overline L}
 {\overline L}> \;\; .
\end{equation}
It was argued{\cite o}
that only operators of the form $<{\cal C} > = <LLLL>$
should contribute as such condensates will have non-trivial flavour
quantum numbers. No rigorous justification was given but this did
allow a systematic study to be undertaken. We note, however, that this
argument breaks down when condensates with the $\bf 144$ alone are
considered. Necessarily, such condensates will be flavour singlets
from the fact that the $\bf 144$ itself is a flavour singlet.

We have seen that higher dimensional models in CSDR constructed on
internal manifolds without $U(1)$ factors are not phenomenologically
acceptable. This means that any gauge invariant structures arising in
acceptable models must be constrained by these factors. Clearly, this
will be important for constructing four fermion operators. Ideally,
fermionic states could be derived from ten dimensional representations
with appropriate $U(1)$ factors such that operators like $<{\cal C}> =
<LLLL>$ could be formed. However, we note that intermediate symmetry
breaking is associated with four fermion condensates of the type
\begin{equation}
<{\cal C}> = <({\bf 144 \times 144}) \times ({\bf 144 \times 144})>
\;\; .
\end{equation}
Not all the $\bf 144$ factors can carry the same additional $U(1)$
charge if this is to be a gauge singlet. But if we have $\bf 144$
representations with differing $U(1)$ quantum numbers then strictly
they belong to inequivalent families. i.e. we would have more than
one such state. So while it may be possible, although it seems
unlikely,
to construct condensates
such as
\begin{equation}
<{\cal C}> = <({\bf 16 \times 16}) \times ({\bf 144 \times 144})>
\;\; ,
\end{equation}
which could include family mixing among $\bf 16$'s with different
$U(1)$ factors,
states of the
form $<{\cal C}> = <LLLL>$ will not in general be gauge singlets. On
the other hand, condensates which can be written as
$<{\cal C}> = <LL{\overline L}{\overline L}>$
can be made to be gauge singlets
without resorting to involved
family mixing prescriptions. Note that this $U(1)$ factor has
arisen before with respect to $SU(5)$ models in CSDR where high colour
condensates were considered for electroweak symmetry breaking
only{\cite p}.
In the light of
previous discussion we will take these $U(1)$ factors as a model
building $constraint$ which binds us to condensates of a particular
type. Such a constraint could clearly not arise in the original four
dimensional approach.

As we have mentioned, higher dimensional models yielding $SO(10)$ in
four dimensions have already been demonstrated to exist. As with the
$S^{6}$ example, many of these models fail to eliminate unwanted
geometrical Higgs fields in the presence of Wilson lines. For example,
an $E_{8}$ model on the manifold $CP^{3} \simeq SU(4)/SU(3)\otimes
U(1)$ will have both $\bf 16$'s and $\bf 144$ fermions arising from the
$\bf 248$ and $\bf 3875$ as well as $\bf 10$'s of Higgs. However, this
manifold has no discrete symmetry in $W${\cite n}
. Consequently, this example
is not viable. The manifold $CP^{2}\otimes S^{2} \simeq SU(3)\otimes
SU(2)/SU(2)\otimes U(1)\otimes U(1)$ is also not suitable since
strictly
$CP^{2}$ cannot support a spinor structure\cite{aa}.
$SO(10)$ models have,
however, been considered on this manifold{\cite l}.
An example constructed on
the manifold $SP(4)\otimes SU(2)/SU(2)\otimes SU(2)\otimes U(1)$
appears promising\cite{l,bb}. Here an $SO(10)$ model with $\bf
16$'s of fermions
emerges with the Higgs fields transforming
as {\bf 10}'s. Unfortunately, the ${\bf 10}(0)$
Higgs state (the number in brackets corresponding to the $U(1)$
charge) corresponding to the vector component $\bf (2,2)$(0) of the
$\bf 6$ of $SO(6)$ transforms to itself under $Z_{2}^{S/R}$
so the $\bf 10$(0) Higgs state
survives. The manifold $Sp(4)/SU(2)\otimes U(1)$ has repeatedly been
applied successfully to $SU(5)$ G.U.T. models{\cite l}.
In particular, a
realistic model on the nonsymmetric version of this manifold has been
previously considered{\cite t}.
However, $SO(10)$ G.U.T. models with this
internal space invariably contain exotic fermions in the $\bf 10$,
and sometimes other representations, of
$SO(10)$. We could attempt to eliminate these by an appropriate
embedding of $Z_{2}^{S/R}$, such as with (3.3), so that
all the fields do not transform evenly under $Z_{2}^{H}$.
Eigenstates of such fermions under
$Z_{2}^{S/R}\oplus Z_{2}^{H}$ may vanish when the Majorana condition is
imposed in ten dimensions{\cite n}. However, as mentioned earlier, the
bidoublet component of the {\bf 10} of $SO(10)$ is a singlet under
this $U(1)$ so such definite eigenstates do not arise.
While exotic fermions have interesting properties within
left-right symmetric models\cite{cc}
, we wish to remain as close as possible to
the minimal anomaly free set (4.2).

We are thus left with two examples, the symmetric manifold
$(SU(2)/U(1))^{3}$ and the nonsymmetric manifold $SU(3)/U(1)\otimes
U(1)$. The interesting advantage of manifolds where $R$ contains more
than one $U(1)$ factor lies in the added freedom to manipulate the
embedding $\Phi :R \rightarrow G$, corresponding to taking new linear
combinations of the $U(1)$ generators. Thus we will present candidate
models on these manifolds and demonstrate that the required fermionic
content arises with Higgs fields of geometrical origin being
eliminated.

\section{Candidate Models}
\subsection{\underline{Example on a symmetric coset space}}

We will consider a $G=E_{8}$ theory on the manifold $M^{4}\otimes
B_{0}$ where $B_{0} = (SU(2)/U(1))^{3}$. A similar model has been
previously investigated but not in the context of the dynamical
symmetry breaking scheme we are considering\cite{bb}.
The $R=(U(1))^{3}$ group is
chosen to be identified with the $(U(1))^{3}$ subgroup of $E_{8}$
appearing in the decomposition
\begin{eqnarray}
E_{8} & \supset & E_{7}\otimes SU(2) \nonumber \\
&\supset & E_{7} \otimes U(1)_{I} \nonumber \\
& \supset & E_{6} \otimes U(1)_{I} \otimes U(1)_{II} \nonumber \\
& \supset & SO(10) \otimes
{\underbrace{U(1)_{I} \otimes U(1)_{II} \otimes U(1)_{III}}_{R}} \;\; .
\end{eqnarray}
We consider fermions to be transforming in the adjoint ${\bf 248}$ and
the ${\bf 3875}$ dimensional representations of $E_{8}$. Note that now
the model is explicitly not supersymmetric. We decompose these
representations under (5.1) by employing the branching rules{\cite w}:
\begin{eqnarray}
E_{8}  & \supset & E_{7} \otimes SU(2) \nonumber \\
{\bf 248} & = &
({\bf 1}, {\bf 3})+({\bf 133}, {\bf 1})+({\bf 56}, {\bf 2}) \nonumber
\\
{\bf 3875} & = & ({\bf 1}, {\bf 1})+({\bf 56}, {\bf 2})+({\bf 133},{\bf
3})+({\bf 1539},{\bf 1})+({\bf 912},{\bf 2}) \;\; .\nonumber \\
SU(2) & \supset & U(1) \nonumber \\
{\bf 2} & = & 1 + -1 \nonumber \\
{\bf 3} & = & 2+0+-2 \;\; . \nonumber
\end{eqnarray}
\begin{eqnarray}
E_{7}  & \supset & E_{6} \otimes U(1) \nonumber \\
{\bf 56} & = & {\bf 1}(3) + {\bf 27}(1)+ {\overline{\bf 27}}(-1)+ {\bf
1}(-3) \nonumber \\
{\bf 133} & = & {\bf 78}(0) + {\bf 1}(0) + {\bf 27}(-2)
+{\overline{\bf 27}}(2) \nonumber \\
{\bf 912} & = & {\bf 78}(3) + {\bf 78}(-3) + {\bf 351}(1) +
{\overline{\bf 351}}(-1) + {\overline{\bf 27}}(-1) +  {\bf 27}(1)
\nonumber
\\
{\bf 1539} & = & {\bf 1}(0) + {\bf 27}(4) + {\overline{\bf 27}}(-4) +
{\bf 27}(-2) + {\overline{\bf 27}}(2) +
{\bf 78}(0) + {\bf 351}(-2) \nonumber \\
 & + &{\overline{\bf 351}}(2) + {\bf
650}(0) \;\; .
\nonumber \\
E_{6} & \supset & SO(10) \otimes U(1) \nonumber \\
{\bf 27} & = & {\bf 1}(4) + {\bf 10}(-2) +{\bf 16}(1) \nonumber \\
{\bf 78} & = & {\bf 1}(0) + {\bf 45}(0) + {\bf 16}(-3) +{\overline{\bf
16}}(3)  \nonumber \\
{\bf 351} & = & {\bf 10}(-2) + {\overline{\bf 16}}(-5) + {\bf 16}(1) +
{\bf 45}(4) + {\bf 120}(-2)+ {\bf 144}(1) \nonumber \\
{\bf 650} & = & {\bf 1}(0) + {\bf 10}(6) + {\bf 10}(-6) + {\bf 16}(-3)
+ {\overline{\bf 16}}(3) + {\bf 45}(0) + {\bf 54}(0)
 \nonumber \\ & + & {\bf 144}(-3)
 +  {\overline{\bf 144}}(3) + {\bf 210}(0) \;\; .
\end{eqnarray}
Thus the four dimensional gauge group will be
\begin{equation}
H=C_{E_{8}}(U(1)^{3})=SO(10)(\otimes U(1)^{3}) \;\; .
\end{equation}
The $R=U(1)^{3}$ content of $(SU(2)/U(1))^{3}$ vector and spinor
are{\cite r}
\begin{eqnarray}
\underline{6} = (2a,0,0) + (0,2b,0) +(0,0,2c) + (-2a,0,0) + (0,-2b,0)
+(0,0,-2c) \nonumber \\
\underline{4} = (a,b,c) + (-a,-b,c) + (-a,b,-c) + (a,-b,-c) \;\;\;\;
\;\;\;\;\;\;\;\;\; .
\end{eqnarray}
Applying the CSDR rules with $a=b=c=1$ we get
\newline
(a) scalar fields transforming as ${\bf 1}(2,0,0)+{\bf 1}(-2,0,0)$ and
\newline
(b) fermions transforming as $3\times {\bf 16}(1,1,1)+{\bf 144}(1,1,1)
+ 3\times {\bf 16}(-1,1,1) + {\bf 144}(-1,1,1)$.

This manifold has a $(Z_{2})^{3}$ symmetry in $W$, where each $Z_{2}$
changes the sign of $a,b$ and $c${\cite n}.
We take $Z_{2}^{S/R} \subset
(Z_{2})^{3}$ and embed this into the $U(1)_{I}\otimes U(1)_{II}$
subgroup of $SO(10)$
appearing in the decomposition
\begin{eqnarray}
SO(10) &\supset & SU(5)\otimes U(1)_{I} \nonumber \\
& \supset & SU(2)_{L}\otimes SU(3)_{C}\otimes U(1)_{I}\otimes
U(1)_{II} \;\; ,
\end{eqnarray}
and in such a way that all the fields transform evenly under
$Z_{2}^{H}$. We choose a solution for the symmetry breaking matrices
$U$, arising from the homomorphism $Z_{2}^{S/R}\rightarrow G$, such
that a maximal number of unbroken generators survive,
resulting in the four dimensional gauge group{\cite e}
\begin{equation}
K=C_{E_{8}}((U(1))^{3}\otimes Z_{2}^{H})=SU(2)\otimes SU(2)\otimes
SU(4)(\otimes (U(1))^{3}) \;\; .
\end{equation}
We could alternatively have chosen the embedding given in (3.3).
Under the action of $Z_{2}^{S/R}\oplus Z_{2}^{H}$ eigenstates of the
scalar fields do not have definite transformation properties under the
four dimensional gauge group and so are eliminated. It is worthwhile
pointing out, however, that gauge singlet scalar fields can be very
useful in phenomenological model building. Recalling the Majorana
condition, the
fermionic fields survive.

We see that we have two sets of fermions transforming as $3\times {\bf
16} + {\bf 144}$ under $SO(10)$. Since we do not wish to introduce any
mixing between these sets we will choose to identify them by an
appropriate choice of discrete symmetry on the complex structure of
the internal space such that
\begin{equation}
Z_{2} : (1,1,1) \leftrightarrow (-1,1,1) \;\; .
\end{equation}

Thus we have realized a model in four dimensions with the appropriate
set of fermions. Note that we need not have necessarily embedded
$Z_{2}^{S/R}$ in $H$ since we have chosen the fields to transform
evenly under $Z_{2}^{H}$. In this case we would have an $SO(10)$ model
in four dimensions with the appropriate fermionic content.

\subsection{\underline{Example on a nonsymmetric coset space}}

We will consider now a $G=E_{7}$ theory on the manifold $M^{4}\otimes
B_{0}$ where $B_{0}= SU(3)/U(1)\otimes U(1)$. The $R=U(1)\otimes U(1)$
group is chosen to be identified with the $U(1)\otimes U(1)$ subgroup
of $E_{7}$ appearing in the decomposition
\begin{eqnarray}
E_{7} & \supset & SO(12)\otimes SU(2) \nonumber \\
& \supset & SO(12)\otimes U(1)_{I} \nonumber \\
& \supset & SO(10)\otimes {\underbrace{
U(1)_{I} \otimes U(1)_{II}}_{R}} \;\; .
\label{decomp}
\end{eqnarray}
We take fermions to be transforming in two adjoint {\bf 133}'s and one
{\bf 1463} dimensional representation of $E_{7}$. These
representations are decomposed under (\ref{decomp}) by employing the
branching rules{\cite w}:
\begin{eqnarray}
E_{7} & \supset & SO(12)\otimes SU(2) \nonumber \\
{\bf 133} &= &({\bf 1},{\bf 3})+({\bf 32}',{\bf 2})+({\bf 66},{\bf 1})
\nonumber \\
{\bf 1463} &= &
({\bf 66},{\bf 1})+({\bf 77},{\bf 3})+({\bf 462},{\bf 1})+
({\bf 352}',{\bf 2}) \;\; .\nonumber \\
SU(2) & \supset & U(1) \nonumber \\
{\bf 2} & = & 1 + -1 \nonumber \\
{\bf 3} & = & 2+0+-2 \;\; . \nonumber \\
SO(12) & \supset & SO(10)\otimes U(1) \nonumber \\
{\bf 32}' & = & {\bf 16}(-1) + {\overline{\bf 16}}(1) \nonumber \\
{\bf 66}& = &
{\bf 1}(0)+{\bf 45}(0)+{\bf 10}(2)+{\bf 10}(-2)
\nonumber \\
{\bf 77} & = &
{\bf 54}(0)+{\bf 10}(2)+{\bf 10}(-2)+{\bf 1}(4)+
{\bf 1}(0)+{\bf 1}(-4) \nonumber \\
{\bf 352}' & = &
{\bf 144}(-1)+{\overline{\bf 144}}(1)+{\overline{\bf 16}}(-3)
+{\overline{\bf 16}}(1)+{\bf 16}(-1)+{\bf 16}(3) \nonumber \\
{\bf 462}& = &
{\bf 126}(2)+{\overline{\bf 126}}(-2)+{\bf 210}(0) \;\; .
\end{eqnarray}
Thus the four dimensional gauge group will be
\begin{equation}
H=C_{E_{7}}(U(1)\otimes U(1))=SO(10)(\otimes U(1)\otimes U(1)) \;\; .
\end{equation}
The $R=U(1)\otimes U(1)$ content of $SU(3)/U(1)\otimes U(1)$ vector
and spinor are{\cite r}
\begin{eqnarray}
{\bf 6} & = & (a,c)+(b,d)+(a+b,c+d) \nonumber \\
& + & (-a,-c)+(-b,-d)+(-a-b,-c-d) \nonumber \\
{\bf 4} & = & (0,0) + (a,c)+(b,d)+(-a-b,-c-d) \;\; .
\end{eqnarray}
We make a particular choice of embedding $\Phi : R \rightarrow G$ by
setting $a=1,c=-1,b=1,d=0$. Applying the CSDR rules we get
\newline
(a) scalar fields transforming as ${\bf 16}(1,-1) + {\overline{\bf
16}}(-1,1)$ and
\newline
(b) fermions transforming as $3\times {\bf 16}(1,-1) + {\bf
144}(1,-1)$.
We have neglected gaugino fermionic fields as they all obtain masses
on the order of the compactification scale from the purely geometrical
term, $V${\cite y}, which appears in the fermionic mass matrix.

This manifold has a $Z_{2}$ discrete symmetry in $W${\cite n}
. Under this the
$R$ decompositions of the vector and spinor transform as
\begin{equation}
Z_{2}^{S/R} =
\left \{
\begin{array}{c}
( 1,  0) \leftrightarrow ( 2, -1) \\
( 1,  -1) \leftrightarrow ( -1,  1) \\
( -1,  0) \leftrightarrow ( -2, 1) \\
\end{array}
\right .
\end{equation}
We can again embed this into the same $U(1)$ subgroup of $SO(10)$ as
before such that all the fields transform evenly. Eigenstates of the
scalar fields under $Z_{2}^{S/R}\oplus Z_{2}^{H}$ do not have definite
transformation properties under the four dimensional gauge group so do
not survive. Recalling the Majorana condition, the fermionic fields do
survive.

Thus again we have realized a model in four dimensions with the
appropriate fermionic content. Note that in both models the
$Z_{2}^{S/R}$ discrete symmetry can substitute the Majorana condition
which we consequently relax.

\section{The Dynamical Symmetry Breaking Scheme}
As with Napoly{\cite o},
we will take all symmetry breaking from the four
dimensional gauge group
down to $SU(3)_{C}\otimes U(1)_{Q}$ to originate from the existence of
$G$-symmetric four fermion condensates. Decomposing $\bf R_{L}\times
R_{L}$ we have{\cite w}
\begin{eqnarray}
{\bf 16}\times {\bf 16} & = & {\bf 10}_{S} + {\bf 120}_{A} + {\bf
126}_{S} \nonumber \\
({\bf 144 \times 144})_{S} & = & {\bf 10} + {\bf 126} + {\overline{\bf
126}} + {\bf 210}' + {\bf 320} + {\bf 1728} \nonumber \\
& + & {\bf 2970} + {\bf 4950} \nonumber \\
{\bf 16} \times {\bf 144} & = & {\bf 10} + {\bf 120} +
{\overline{\bf 126}} +
{\bf 320} + {\bf 1728} \;\; .
\end{eqnarray}
Since we are considering condensates of the form $<{\cal
C}>=<LL{\overline L}{\overline L}>$, the {\bf 4950} dimensional
representation will contribute even though it is complex. The symmetry
breaking pattern can proceed along two possible directions depending
on whether we (i) embed $Z_{2}^{S/R}$ into the four dimensional gauge
group or (ii) choose not to embed $Z_{2}^{S/R}$. The resulting breaking
schemes then have the form
\newline
\newcounter{cms}
\setlength{\unitlength}{1mm}
\begin{picture}(200,60)(0,-1)
\put(25,50){\makebox(0,0){$SO(10)$}}
\put(6,50){\makebox(0,0){$G$}}
\put(1,50){\makebox(0,0){(i)}}
\put(67,50){\makebox(0,0){$SU(2)_{L}\otimes SU(2)_{R}\otimes SU(4)$}}
\put(61,25){\makebox(0,0){$SU(2)_{L}\otimes SU(3)_{C}\otimes
U(1)_{Y}$}}
\put(59,0){\makebox(0,0){$SU(3)_{C}\otimes U(1)_{Q}$}}
\put(13,53){\makebox(0,0){$M_{Pl}$}}
\put(37,53){\makebox(0,0){$M_{Pl}$}}
\put(64,38){\makebox(0,0){$M_{I}$}}
\put(65,13){\makebox(0,0){$M_{II}$}}
\thicklines
\put(8,50){\vector(1,0){10}}
\put(32,50){\vector(1,0){10}}
\put(60,45){\vector(0,-1){15}}
\put(60,20){\vector(0,-1){15}}

\put(125,50){\makebox(0,0){$SO(10)$}}
\put(106,50){\makebox(0,0){$G$}}
\put(100,50){\makebox(0,0){(ii)}}
\put(128,25){\makebox(0,0){$SU(2)_{L}\otimes SU(3)_{C}\otimes
U(1)_{Y}$}}
\put(126,0){\makebox(0,0){$SU(3)_{C}\otimes U(1)_{Q}$}}
\put(113,53){\makebox(0,0){$M_{Pl}$}}
\put(131,38){\makebox(0,0){$M_{I}$}}
\put(132,13){\makebox(0,0){$M_{II}$}}
\thicklines
\put(108,50){\vector(1,0){10}}
\put(127,45){\vector(0,-1){15}}
\put(127,20){\vector(0,-1){15}}
\end{picture}
\newline
where $M_{Pl}$ is the Planck scale and $M_{I}$ and $M_{II}$ are mass
scales characterizing each level of symmetry breaking. It is
worthwhile emphasising that all the symmetry breaking is induced
either by condensates or topologically. This is true even in case (i)
where we employ Wilson lines. It has been pointed out that Wilson
lines are similar to ordinary Higgs fields transforming in the adjoint
representation of $H${\cite b}
. However, it is not difficult to show that it is
possible, by a more exotic embedding of $R$ into $G$, to achieve the
breaking pattern $G\rightarrow SU(2)_{L}\otimes SU(2)_{R}\otimes
SU(4)$ directly by the CSDR mechanism alone. It is not clear that a
similar result holds for $SU(5)$ models. Thus, at the expense of a
more involved procedure, we could have arrived at similar conclusions
without Wilson lines. We can, therefore, maintain the model building
prescription without introducing Higgs fields outside those formed by
condensates.

We will assume that only the {\bf 144} dimensional representation is
involved in forming Higgs condensates at the scales $M_{I}$ and
$M_{II}$. In this way we will be able to give these exotic fermions
large masses relative to the {\bf 16}'s. At the scale $M_{I}$, then,
we need only consider the {\bf 126}, $\overline{\bf 126}$, {\bf 1728},
{\bf 2970} and {\bf 4950} representations in forming effective Higgs
fields as these contain a $SU(2)_{L}\otimes SU(3)_{C}\otimes U(1)_{Y}$
singlet{\cite w}
. We can therefore consider the condensates, in the notation of
Napoly{\cite o}:
\begin{eqnarray}
<{\cal C}'_{I}> & = & <[({\bf 144\times 144})_{{\bf 126}}
\times ({\overline{\bf
144}}\times {\overline{\bf 144}})_{\overline{\bf 126}}]_{{\bf 1}}>
\sim (M_{I}')^{6} \nonumber \\
<{\cal C}''_{I}> & = & <[({\bf 144\times 144})_{{\bf 1728}}
\times ({\overline{\bf
144}}\times {\overline{\bf 144}})_{\bf 1728}]_{{\bf 1}}>
\sim (M_{I}'')^{6} \nonumber \\
<{\cal C}'''_{I}> & = & <[({\bf 144\times 144})_{{\bf 2970}}
\times ({\overline{\bf
144}}\times {\overline{\bf 144}})_{\bf 2970}]_{{\bf 1}}>
\sim (M_{I}''')^{6} \nonumber \\
<{\cal C}''''_{I}> & = & <[({\bf 144\times 144})_{{\bf 4950}}
\times ({\overline{\bf
144}}\times {\overline{\bf 144}})_{\overline{\bf 4950}}]_{{\bf 1}}>
\sim (M_{I}'''')^{6} \;\; .
\end{eqnarray}
The energy scales corresponding to each condensate are assumed to be
approximately equal. Unfortunately, a satisfactory topological
mechanism to inhibit proton decay has yet to be found within this
higher dimensional scenario{\cite u}. We are therefore moved to set the
scale $M_{I}$ greater than about $10^{14}$Gev so that
\begin{equation}
M_{I}' \simeq M_{I}'' \simeq M_{I}''' \simeq M_{I}'''' \geq 10^{14}
{\rm Gev} \;\; .
\end{equation}
If we choose to realize breaking scheme (i), however, all the baryon
number violating gauge fields will have been made superheavy at the
compactification scale. In this case then we could set the scale
$M_{I}$ to be significantly lower ($10^{6}-10^{7}$Gev for a light
$W_{R}$ model).
Note that we could also have included the condensate $<[({\bf 144
\times 144})_{\overline{\bf 126}} \times ({\overline{\bf 144}}\times
{\overline{\bf 144}})_{\bf 126}]_{\bf 1}>$. With no compelling reason
to the contrary, we will simply take this to be also characterized by
the energy scale $M_{I}'$.

An appropriate choice of four fermion condensates at the scale $M_{II}$
corresponds to effective Higgs fields transforming as {\bf 10}, ${\bf
210}'$ and {\bf 320} which all contain an $SU(2)_{L}$ doublet. We
therefore have the condensates
\begin{eqnarray}
<{\cal C}'_{II}> & = & <[({\bf 144\times 144})_{{\bf 10}}
\times ({\overline{\bf
144}}\times {\overline{\bf 144}})_{\bf 10}]_{{\bf 1}}>
\sim (M_{II}')^{6} \nonumber \\
<{\cal C}''_{II}> & = & <[({\bf 144\times 144})_{{\bf 210}'}
\times ({\overline{\bf
144}}\times {\overline{\bf 144}})_{{\bf 210}'}]_{{\bf 1}}>
\sim (M_{II}'')^{6} \nonumber \\
<{\cal C}'''_{II}> & = & <[({\bf 144\times 144})_{{\bf 320}}
\times ({\overline{\bf
144}}\times {\overline{\bf 144}})_{{\bf 320}}]_{{\bf 1}}>
\sim (M_{II}''')^{6} \;\; .
\end{eqnarray}
It is known that this symmetry breaking occurs at around $10^{2}$Gev
so that
\begin{equation}
M_{II}'\simeq M_{II}''\simeq M_{II}''' \simeq 10^{2} {\rm Gev} \;\; .
\end{equation}
As has been pointed out{\cite o}
, the ${\bf 210}'$ and {\bf 320} also contain
Higgs representations, transforming as $({\bf 4},{\bf 1})(1)+ ({\bf 4}
 ,{\bf 1})(-1)$ under $SU(2)_{L}\otimes SU(3)_{C}\otimes U(1)_{Y}$,
that can induce breaking to $SU(3)_{C}\otimes U(1)_{Q}$. However,
these representations always appear with doublets, while doublets can
appear alone, so that it seems reasonable to postulate that
symmetry breaking occurs by composite Higgs doublets only.

\section{The Fermion Mass Spectrum and Chiral Symmetry Breaking}

We propose that the quark and lepton bare masses
originate from the condensates
\begin{eqnarray}
<{\cal C}_{f}> & = & <[({\bf 16\times 16})_{{\bf 10}}
\times ({\overline{\bf
144}}\times {\overline{\bf 144}})_{\bf 10}]_{{\bf 1}}>
\neq 0 \nonumber \\
<{\cal C}'_{f}> & = & <[({\bf 16\times 16})_{{\bf 126}}
\times ({\overline{\bf
144}}\times {\overline{\bf 144}})_{\overline{\bf 126}}]_{{\bf 1}}>
\neq 0 \;\; ,
\end{eqnarray}
while chiral symmetry breaking in the quark sector is associated with
\begin{equation}
<{\cal C}_{c}>  =  <[({\bf 16\times 16})_{{\bf 10}}
\times ({\overline{\bf
16}}\times {\overline{\bf 16}})_{\bf 10}]_{{\bf 1}}>
\sim (M_{c})^{6} \;\; ,
\end{equation}
where $M_{c}$ is the symmetry breaking scale such that $M_{c} <<
M_{II}$. Note that this approach divorces the chiral symmetry breaking
from the electroweak breaking at $M_{II}$ which the additional $U(1)$
factors in $H$ obstruct in $SU(5)$ models previously considered under
CSDR which utilize dynamical symmetry breaking{\cite n}.
The condensate ${\cal C}_{f}$ couples fermions to the {\bf 10}
composite Higgs as in a Yukawa coupling. This mass scale is suggested
to occur at $M_{c} \simeq 10^{-1} {\rm Gev}$ at which the {\bf 144}
fermions are decoupled. A large Majorana mass for the antineutrino
comes from coupling to the $({\overline{\bf 144}}\times
{\overline{\bf 144}})_{{\overline{\bf 126}}}$ composite Higgs at the
$M_{I}$ scale.

The {\bf 144} fermions attain masses at the scales $M_{I}$ and
$M_{II}$. The effective $SU(2)_{L}$ symmetry above $M_{II}$ protects
the {\bf 144} components associated with symmetry breaking at this
scale so that their masses are not larger than $M_{II}$. This gives
rise to a Tev scale hadronic spectroscopy as well as charged heavy
leptons and neutrinos. Most interestingly, a charge two heavy lepton
emerges.

It is important to
note that, unlike the purely phenomenological model, we do not have an
exact $U(3)$ family symmetry. Originating from different
representations of $G$, a reduced family symmetry at best can exist.
Indeed, it may be possible to find a model in which all the {\bf 16}'s
originate in different representations in ten dimensions. This would
completely eliminate the existence of exactly massless Goldstone
bosons arising from breaking family symmetry. Interestingly, we can
still eliminate these fields by demanding that they be odd under the
$Z_{2}^{S/R}$ discrete symmetry.

\section{Conclusion}
We have presented models which realize an anomaly free set of fermions
necessary to yield realistic low energy theories by the formation of
Higgs fields from fermionic condensates. As well as providing an
origin for this approach within phenomenological models, we now have a
means by which the desert region may be filled in higher dimensional
theories. This is a major problem with $SU(5)$ type models derived
from CSDR. The large gap separating the compactification scale,
usually taken as the Planck scale, and the electroweak scale is
unnatural, yielding apparently no new phenomenology in this region.
While, as with the Wetterich
model{\cite h}, we have set the symmetry breaking
scales by hand we have demonstrated that these symmetry breaking
structures can be associated with fields {\it derived} from higher
dimensional models. Furthermore, the CSDR scheme has provided an
explicit model building constraint for the form of condensates for
which only heuristic arguments could be previously used. It is
compelling also to speculate that a non-trivial topological signature
may arise such that these condensates could have well defined
expectation values.

In the absence of manifolds with the appropriate holonomy or
compelling low energy supersymmetric symmetry breaking schemes, the
introduction of dynamical symmetry breaking provides a consistent
approach to higher dimensional $SO(10)$ unified models. Indeed it has
been noted in left-right symmetric models that conclusions are
unaltered if Higgs fields are replaced by fermionic bilinears\cite{dd}.
Outstanding
questions, such as the hierarchy of scales may yet yield to a more
exotic geometrical approach, while interesting exotic heavy fermions
may provide the experimental signature for these symmetry breaking
mechanisims.

\section{Acknowledgements}
It is a pleasure to thank J. Choi for helpful discussions. BEH
would also like to thank R. Delbourgo and the
members of the theoretical physics group, University of Tasmania, for
their hospitality over the period during which some of this
work was completed. BEH is supported by an Australian Postgraduate
Research Award.

\end{document}